\documentclass[aps,prc,showpacs,twocolumn]{revtex4}
\usepackage{epsfig}
\begin{document}
\title{Charged three-body system with arbitrary masses near conformal invariance}
\author{A. Delfino$^1$, T. Frederico$^2$, and Lauro Tomio$^{3,4}$}
\affiliation{$^1$$^d$Instituto de F\'\i sica, Universidade Federal Fluminense,
24210-900, Niter\'oi, RJ, Brazil.\\
$^2$Departamento de F\'\i sica, Instituto
Tecnol\'ogico de Aeron\'autica, 12228-900 S\~ao Jos\'e dos Campos, Brazil.\\
$^3$Instituto de F\'\i sica Te\'orica, UNESP - Universidade
Estadual Paulista, 01140-070, S\~{a}o Paulo, Brazil.\\
$^4$Centro Brasileiro de Pesquisas F\'\i sicas,
Rua Dr. Xavier Sigaud, 150, 22290-180, Rio de Janeiro, Brazil.}
\date{\today}
\begin{abstract}
Within an adiabatic approximation to the three-body Coulomb system,
we study the strength of the leading order conformaly invariant
attractive dipole interaction produced when a slow charged particle
$q_3$ (with mass $m_3$) is captured by the first excited state of a
dimer [with individual masses and charges $(m_1,q_1$) and
($m_2,q_2=-q_1$)]. The approach leads to a universal mass-charge
critical condition for the existence of three-body level
condensation, $\displaystyle{\left(m_1^{-1}+m_2^{-1}\right)}
/{\left[(m_1+m_2)^{-1}+m_3^{-1}\right]}>\left|{q_1}/(24\;
q_3)\right|$, as well as the ratio between the geometrically scaled
energy levels. The resulting expressions can be relevant in the
analysis of recent experimental setups with charged three-body
systems, such as the interactions of excitons, or other
matter-antimatter dimers, with a slow charged particle.
\end{abstract}
\pacs{36.10.-k,03.65.Ge,21.45.-v,36.10.Gv}
%03.65.Ge Solutions of wave equations: bound states
%36.10.-k Exotic atoms and molecules
%36.10.Gv Mesonic, Hyperonic and antiprotonic atoms and molecules
%21.45.-v Few-body systems
%21.10.Dr Binding energies and masses
\maketitle

\section{Introduction}
In view of the actual experimental possibilities, we recall some general characteristics
of three-body charged systems with arbitrary masses and two different charges, in order
to derive the charge-mass dependence of the leading order strength of the attractive
dipole interaction produced by a bound two-body subsystem (with individual charges $q_1$
and $q_2=-q_1$) in the third particle with charge $q_3$.
We start our investigation by considering the old and well-known case of the interaction
of a positronium ($Ps$, a bound-state of an electron $e^-\equiv -e$ and a positron $e^+\equiv +e$,
where $e$ is the absolute value of the electron charge) with a spectator electron.
This is the ionized negative positronium ($Ps^-$).
Early calculations, by Wheeler in 1946~\cite{wheeler}, have already predicted
a bound state for such system, confirmed by Mills in 1981~\cite{mills}.

Further investigations on the properties of $e^-e^+e^-$
system~\cite{botero86,rost92,papp05,ps93-08}, as well as on other
Coulombic three-body systems, since 1960's up to recent
years~\cite{kolos60,pekeiris62,3coul64-08,lin95,pr-richard05},
have also been motivated by the increasing interest in
matter-antimatter interactions~\cite{schultz91}. Up to 1995, the
theoretical and experimental advances in understanding Coulombic
three-body systems and matter-antimatter interaction can be found,
respectively, in two reviews: \cite{lin95} and \cite{schultz91}.
As emphasized in \cite{schultz91}, small number of leptons,
electrons, muons and their antiparticles, are important to test
fundamental theories of quantum electrodynamics; and systems with
small number of protons and antiprotons can also provide relevant
tests of the strong interaction. The actual interest on the
properties of few-body charged systems is evidenced by the recent
report on the production of a molecular di-positronium
$Ps_2$~\cite{nature07}. For a recent review, particularly
concerned on the stability of quantum charged few-body systems,
see \cite{pr-richard05}. In Ref.~\cite{rost92}, Rost and Wintgen
have explored and classified the $Ps^-$ dynamics considering the
observation that it has a molecular structure similar to the
ionized hydrogen molecule $H_2^+$ [$(pe^-)p$]. They have also
reported the existence of a $^1S$ resonance pattern unknown in
three-body Coulomb systems. Such results, combined with results
obtained for the hydrogen ion $H^-$ [$(pe^-)e^-$], lead them to
suggest the existence of a similar resonance spectrum for all
$ABA$ Coulomb systems with charge and mass ratios such that
$|q_A/q_B|=1$ and $m_A/m_B\ge 1$, respectively. More recently, by
considering a Molecular Adiabatic (MA) treatment for the $Ps^-$,
it was also reported an accumulation of three-body resonances {\it
above} the two-body threshold~\cite{papp05}.

Surprisingly, besides the number of recent studies concerned with few-charged
quantum systems~\cite{pr-richard05}, and the recognized relevance of a mass-charge
universal relation in view of actual experimental facilities~\cite{rost92},
a particular straightforward condition relating charges and masses for a general
three-body system (with arbitrary masses) is still missing in the literature.
In the case of exotic molecular three-body systems, where the subsystem is
hydrogen-like (with $e^-$ replaced by $\mu^-$ or $\pi^-$), it was
shown in \cite{kilic04} that a relation for the structure of the spectrum can be
obtained in the frame of Born-Oppenheimer approximation, following spectral
properties of long-ranged $1/R^2$ interactions~\cite{landau},
which are known to be conformally invariant~\cite{camblong01,braaten06}.

The study of resonance patterns, which can occur in few-body
interactions, became very relevant in trapped ultracold atom
experiments, as the presence of several resonances at
experimentally accessible magnetic fields can allow precise
tunning of atom-ion interaction~\cite{idziaszek09}. Theoretical
predictions such as the increasing number of three-body bound
states when the two-body scattering length goes to infinity, known
as Efimov effect~\cite{Efimov}, can actually be checked
experimentally in ultracold atomic laboratories~\cite{grimm}. The
spectrum of Efimov states, exhibiting a geometrical scaling, is
generated by an attractive effective potential proportional to
$1/R^2$, where $R$ is the distance of one of the particles to the
center-of-mass of the remaining pair, considering short-range
two-body interactions~\cite{Efimov,fonseca} (On the scaling
mechanism and conformal invariance behind this effect, see
\cite{braaten06,camblong01,scaling}). As it will be shown, the
long-ranged Coulombic interactions, for certain configurations of
three charged particles, can exhibit the same kind of effective
interaction.

For recent relevant applications in ultracold laboratories, of a
study with three charged systems, we can mention the possibility
of exciton~\cite{butov02,tan2005} or
positronium~\cite{platzman94,cassidy05} condensed gas interacting
with a charged particle. In case of excitons (electron-hole bound
pair in a semiconductor), the electron and hole in the interacting
pair can acquire effective individual masses distinct from the
free electron mass~\cite{tan2005}. When interacting with a slow
charged particle, a charge-mass dependent resonance pattern should
emerge. So, well-based simple charge-mass conditions can be very
helpful to analyze the relation between effective masses and the
observed spectrum.

{Motivated by the above discussion we show examples in
atomic physics of charged three-body systems in which appears an
effective long-range $1/R^2$ potential, where the strength is
modulated by the arbitrary individual masses and charges}, restricted to the
condition that the bound subsystem is neutral. A robust
mass-charge critical condition is derived for the weakly-bound
three-body spectrum {\it below} the state $n=2$ of the subsystem,
in case of arbitrary masses. Here, we should observe that the
degeneracy between the $2S$ and $2P$ levels of the chargeless
two-body subsystem is broken by the dipole potential. When it is
attractive it rises to the spectrum {\it below}\cite{rost92} that
we consider; the repulsive part is responsible for a set of
resonances {\it above} the $2S-2P$ state~\cite{papp05}. The
approach is strictly valid for the cases where the dimer is made
up with {\it point-like} charged particles. However, it can also
be taken as a first approach when the charged particles are more
complex objects.

{The basic equations of our formalism, in the adiabatic approximation, 
are given in the next section, which lead to the effective $1/R^2$ long-range
interaction and a spectrum geometrically scaled. In section III, after
analyzing the range of values for the strength of the effective interaction 
in terms of the particle masses, we illustrate the main results with 
examples. A general discussion on the applicability of the
present approach and a summary of our conclusions are provided 
in section IV.}

\section{Formalism}

In this section, we present the basic formalism for a general charged three-body
system, where the masses ($m_1, m_2, m_3$) can be arbitrary, and the charges are such
that we have a bound chargeless two-body subsystem ($q_2=-q_1$)
interacting with a slow charged particle $q_3$
(the charges are in units of the absolute value of the electron charge $|e|$).
The system Hamiltonian is given by
\begin{eqnarray}
H &=& -\frac{\hbar^2}{2\mu_{R}}\nabla_R^2 -\frac{\hbar^2}{2\mu_{r}}\nabla_r^2
+ V(\vec{r},\vec{R}),\label{eq01}\\
V(\vec{r},\vec{R})&\equiv&
\frac{q_1q_2}{r} + \frac{q_2q_3}{|{\frac{\mu_{r}}{m_2}{\vec r}
+\vec R}|} + \frac{q_3q_1}{|{\frac{\mu_{r}}{m_1}{\vec r}-\vec R}|},
\label{eq02}\end{eqnarray}
where $\vec r$ is the distance vector between the charges $q_1$ and $q_2$
and $\vec R$ the distance vector from the center-of-mass of the subsystem to the
third (spectator) particle.
Here, we define $\mu_r$ as the reduced mass of the subsystem $(q_1q_2)$,
$\mu_r\equiv\mu_{12}\equiv (m_1^{-1}+m_2^{-1})^{-1}$; and $\mu_R$ the
corresponding reduced mass of the subsystem $(q_1 q_2)$ interacting with the
third particle:
$\mu_R\equiv \mu_{(12)3}\equiv [(m_1+m_2)^{-1}+m_3^{-1}]^{-1}$.

\subsection{Adiabatic approximation for three charges} 
The adiabatic approximation implies in solving a coupled equation for
the total wave-function $\Psi(\vec R,\vec r) = \psi(\vec R,\vec
r)\Phi(\vec R)$, such that we first solve the Schr\"odinger equation
in the variable $\vec{r}$ for the wave-function
$\psi(\vec{R},\vec{r})$, using $\vec{R}$ as a parameter, with energy
solution $U(\vec R)$:
\begin{equation}
\left(
-\frac{\hbar^2}{2\mu_{r}}\nabla_r^2 + V(\vec{r},\vec{R})
\right)\psi(\vec{R},\vec{r}) = U(\vec R) \psi(\vec{R},\vec{r}).
\label{eq03}
\end{equation}
Next, the total energy $E$ is given by
\begin{eqnarray}
\left(
-\frac{\hbar^2}{2\mu_{R}}\nabla_R^2 + U(\vec R)
\right)\Phi(\vec{R}) &=& {E} \Phi(\vec{R}).
\label{eq04}\end{eqnarray}
The separation of the Schr\"odinger equation by means of the
adiabatic approximation is justified in case
the energy of the relative motion of the center-of-mass of
the two-body subsystem, with respect to the third particle, is small
in comparison with the subsystem binding energy.

Here, the main interest is the behavior of $U\equiv U(\vec R)$
as $R$ goes to infinity. The case where the third particle
interacts with the $(q_1q_2)-$subsystem.
Using the expansion for the Coulomb potential, with
$|\vec R|>>|\vec r|$ and
$P_l$ the usual Legendre polynomials of order $l$,
we have
{\small
\begin{eqnarray}
U \psi(\vec{R},\vec{r})\hspace{-0.2cm}&=&\hspace{-0.2cm}
\left[
-\frac{\hbar^2}{2\mu_{r}}\nabla_r^2 + \frac{q_1q_2}{r}+
\frac{q_3}{R}
\sum_{l=0}\left\{q_1\left(\frac{-\mu_rr}{m_1R}\right)^{l}
\right.\right. \nonumber \\ &+&\left.\left.
q_2\left(\frac{\mu_rr}{m_2R}\right)^{l}
\right\}
P_l(\hat{r}.\hat{R})
\right]\psi(\vec{R},\vec{r})
.\label{eq05}
\end{eqnarray}
}
Assuming a neutral subsystem ($q_1=-q_2$), the leading order term
(conformally invariant) in the asymptotic expansion $R\rightarrow\infty$ is given by
{\small \begin{equation}
{U}\psi(\vec{R},\vec{r})=
\left[
-\frac{\hbar^2}{2\mu_{r}}\nabla_r^2 - \frac{q_1^2}{r}
- \frac{q_1q_3 r}{R^2}P_1(\cos\theta)
\right]\psi(\vec{R},\vec{r}),
\label{eq06}\end{equation}
}
where $\theta$ is the angle between the vectors $\vec r$ and $\vec R$.
The next-to-leading order contribution to the adiabatic potential is
 ${\cal O}(R^{-4})P_3(\cos\theta)$ if $m_1=m_2$, otherwise a term like
${\cal O}(R^{-3})P_2(\cos\theta)$ appears, which however does not contribute
due to parity conservation. Note that the higher order terms in Eq.(\ref{eq05}) break
the conformal invariance.

The interaction between the third particle and the subsystem $(q_1q_2)$
at large distances is dominated by a dipole potential, which breaks the degenerated
character of the opposite parity states.
According to Ref.~\cite{botero86}, there is an accumulation of resonances
below the threshold of the $n=2$ excited state of the positronium negative ion,
due to an attractive dipole interaction, which appears as the leading asymptotic
term of the potential at large hyperspherical radius. Therefore, it is reasonable
to assume that for large values of $R$ the  first low lying levels of the
$(q_1q_2)-$subsystem, i.e., the first excited $S$ and $P$ states are weakly
perturbed by the dipole interaction and the degeneracy is broken.
For $R\rightarrow\infty$, the distorted wave function of the $(q_1q_2)-$subsystem
is a linear combination of atomic orbitals (LCAO), composed by
$\psi(\vec{R},\vec{r}) = a(R) \phi_{(20)}(r)Y_{00}(\hat{r})
+ b(R) \phi_{(21)}(r)Y_{10}(\hat{r})$,
where $Y_{lm}(\hat{r})$ (with $lm=00, 10$) are the usual spherical harmonic function and
$\phi_{nl}$ (with $nl=20, 21$) are the two-body hydrogenic like wave functions for the
charges $q_1$ and $-q_1$.
Inserting $\psi(\vec{R},\vec{r})$ in Eq.~(\ref{eq06}), we obtain
 \begin{eqnarray}
 U(R) &=& - E_1  \pm \frac{q_1q_3}{R^2}|K|,
  \label{eq07} \\
 K&\equiv&\int d^3r \phi^*_{20}(r) Y^*_{00}(\hat r)
 r P_1(\cos\theta )\phi_{21}(r) Y_{10}(\hat r)
\nonumber\\
&=& -3\left({\hbar^2}/{(\mu_{r} q_1^2)}\right).
 \label{eq08}\end{eqnarray}
where $E_1$ is the energy
of the first excited state of the subsystem $(q_1q_2)$. {The
two possible signs in (\ref{eq07}) come from the diagonalization of the
eigenvalue equation (\ref{eq06}) in the 2S-2P subspace.}

From Eqs.~(\ref{eq07}), (\ref{eq08}) and (\ref{eq04}), we obtain the corresponding
equation for the third particle motion related to the
center-of-mass of the subsystem:
\begin{equation}
\left( - \frac{\hbar^2 \nabla^2_R}{2\mu_{R}}
\pm {3\frac{\hbar^2 q_3}{\mu_{r} q_1}}\frac{1}{R^2} \right)
\Phi = (E+E_1)\Phi\equiv
\frac{\hbar^2}{2\mu_{R}}
{\cal E}\Phi.
\label{eq09}
\end{equation}
Here we note how the degeneracy between the $2S$ and $2P$ levels
 of the chargeless two-body subsystem is broken by the dipole potential.
The motion of the wave function $\Phi\equiv\Phi(R)$ is given by an attractive
type potential.
{The effect discussed in Ref.~\cite{rost92} of level
condensation near the $N-th$ state of positronium is
analogous to this situation, as the long-range attractive
potential $1/R^2$ is responsible for forming such resonances.}
The other solution to $U(R)$
gives a long-range repulsion which means that, asymptotically, no
bound states can be formed. This repulsive $1/R^2$ interaction is
responsible for the set of resonances above the $2S-2P$
states~\cite{papp05}. Focusing our approach in the weakly-bound
states of Eq.~(\ref{eq09}), we end up with an attractive dipole
interaction, $\lambda/R^2$ , generated by the particles with
masses $m_1$ and $m_2$, where the dimensionless $\lambda$ is given
by
\begin{eqnarray}
\lambda\,&=& 6\left|\frac{q_3}{q_1}\right|\frac{\mu_R}{\mu_r}=
 6\left|\frac{q_3}{q_1}\right|
 \frac{(m_1^{-1}+m_2^{-1})}{(m_1+m_2)^{-1}+m_3^{-1}}.
\label{eq10}\end{eqnarray} Now, to solve Eq.~(\ref{eq09}), we
employ a { well-known result derived for attractive dipole
potential (see \cite{landau})}, which implies in a spectrum of
infinite weakly-bound levels ${\cal {E}}_\nu$ below $E=-E_1$, if
$\lambda\ge 1/4$; otherwise, no other bound state. So, the following
condition emerges for the existence of such spectrum:
\begin{equation}
 \frac{(m_1^{-1}+m_2^{-1})}{(m_1+m_2)^{-1}+m_3^{-1}}\ge \frac{1}{24}
 \left|\frac{q_1}{q_3}\right|.
 \label{eq11}\end{equation}
In order to examine the condition suggested in \cite{rost92} for
the occurrence of level condensation, we restrict the above to an
$ABA$ system with $|{q_B}|=|{q_A}|$: If $m_3=m_1=m_A$ and
$m_2=m_B$, we can easily verify that in all the cases we will have
$\lambda\ge 1/4$, implying in a level spectrum below the state
$n=2$ of the subsystem $(AB)$. The mass ratio $m_A/m_B$ in this
case can only control the level spacing of the spectrum. However,
if the bound subsystem is of a particle and antiparticle
($m_A\equiv m_1=m_2$), with the
captured particle having mass $m_B$, the condition $\lambda\ge
1/4$ {will give us $m_A/m_B \le 47.5$.
This can be exemplified with the possible configurations
of three-body systems with protons, electrons and their antiparticles,
such as $(e^-e^+)p$, $(e^-e^+)\bar{p}$ and $(\bar{p}p)e^\pm$,
where level spectrum is expected to occur only for
$(e^-e^+)p$ and $(e^-e^+)\bar{p}$.}

\subsection{Geometrical scaling}

Equation (\ref{eq11}) gives us a quite general relation which
allows infinite number of energy levels { that scale
geometrically} in most of three-body systems involving the usual
stable charged particles and their anti-particles. The number of
states are very dense when $\lambda>> 1/4$, as the ratio between
{ the levels is shown to be given by~\cite{kilic04,landau}
$
{\cal {E}}_{\nu-1}/{\cal
{E}}_\nu = \exp(2\pi/\sqrt{\lambda-1/4}),
$
implying in the geometrically scaled energy levels
\begin{equation}
{\cal {E}}_{\nu} = {\cal {E}}_{0} \exp(-2\nu\pi/\sqrt{\lambda-1/4}),
\label{eq12}
\end{equation}
where ${\cal {E}}_{0}$ is a reference energy, solution of Eq.~(\ref{eq09}),
determined by non-adiabatic effects. Observe that Eq.~(\ref{eq09}) is not well
defined for small values of $R$ (of about few Bohr radii), leading to the
collapse of the reference ground-state energy~\cite{landau}.
Therefore, the dipole potential must be modified at some short distance,
which in our case corresponds to a region where non-adiabatic effects start
to be relevant.}

Two other important remarks related to Eqs.~(\ref{eq10})-(\ref{eq11}):
(i) The strength of the dipole interaction, as well as the
corresponding level ratio ${\cal {E}}_{\nu-1}/{\cal {E}}_\nu $
(when the occurrence of infinite number of levels is possible),
depend only on the ratio of reduced masses $\mu_{(12)3}/\mu_{12}$
and ratio of the charges $|q_3/q_1|$, implying that systems such
as $(e^-e^+)e^\pm$, $(K^-K^+)K^\pm$, or $(p\bar p )p$ have the
same $\lambda$ and energy ratios ${\cal {E}}_{\nu-1}/{\cal {E}}_\nu $.
(ii) Another relevant characteristic of (\ref{eq10}) is its dependence
on the specific three-body configuration and identification of the
spectator particle: Let us consider two possible configurations
for the same three-body system, where  $m_3<<m_1$ and  $m_3<<m_2$
and $q_3=q_2=-q_1$. With $\lambda\equiv\lambda_{(ij)k}$ for the
configuration $(m_im_j)m_k$, we obtain $
\lambda_{(12)3}\lambda_{(13)2}\approx {36}.$ An obvious example is
that of $(pe^-)\bar{p}$ or $(\bar{p}e^+){p}$, where
$\lambda\approx 3(m_p/m_e)>>1/4$ should lead to a dense level
spectrum below the $n=2$ state of the subsystem. For the
counterpart configuration, $(p\bar{p})e^\pm$, $\lambda =
12(m_e/m_p)<<1/4$,
{ a similar spectrum is not expected to occur.}

\section{Results and examples}

\begin{figure}[tbh!]
\vspace{-3.8cm}
\centerline{\epsfig{figure=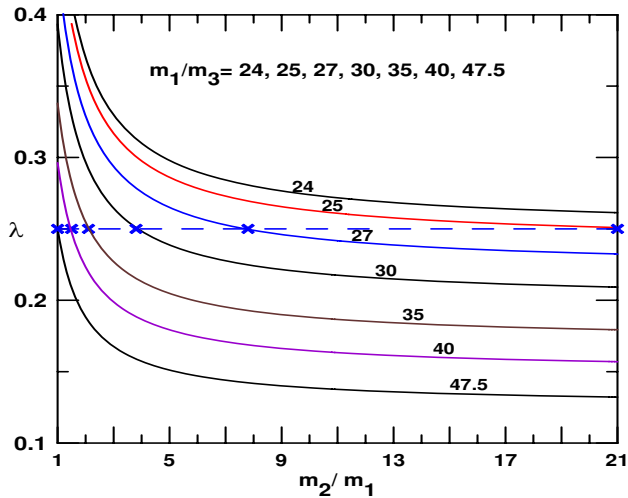,width=12cm,height=13cm}}
\vspace{-3.3cm}
\caption{Behavior of $\lambda$ in terms of the ratio $m_2/m_1$.
Each plot refers to a different ratio $m_1/m_3$,
as given inside the figure. $\lambda>1/4$ is satisfied for all values
of $m_1 < 24 m_3$; and $\lambda<1/4$ for all values of $m_1 > 47.5 m_3$.
} \label{lambda}
\end{figure}

Figure~\ref{lambda} illustrates our conclusions for the values of $\lambda$ as
a function of the mass ratios $m_2/m_1$ and $m_3/m_1$, taking $m_2\ge m_1$  and
$|q_3|=|q_1|$. The critical value of $\lambda=1/4$ is shown by the dashed
straight line.
As observed, only in the cases that $24 < m_1/m_3 < 47.5 $
it is possible to reach the value $\lambda=1/4$ for some ratios of $m_2/m_1$.
There is no infinite number of levels in case that $m_1/m_3 > 47.5$,
independently of $m_2/m_1$. When $m_1/m_3 = 47.5$, the only solution is
$m_2=m_1$; and, when $m_1/m_3 \le 24$, $\lambda >1/4$ for all choices of
$m_2$ and $m_1$.

A few examples of three-body charged systems, with the
corresponding values of $\lambda$ and level ratios ${\cal
{E}}_{\nu-1}/{\cal {E}}_\nu$ are given in Table I. As verified,
{level condensation below the $n=2$ state of the subsystem
(shown in the first two lines of the table) is not possible }
when the mass of the captured particle is much smaller than the
lighter particle of the subsystem. In this table we also present a
few results motivated by the actual interest in matter-antimatter
interaction, and also motivated by experiments with exotic mesonic
atoms [where protons($p$), deuterons($d$) or tritons($t$) are
combined with muons($\mu$), kaons($K$) or pions($\pi$)]. Our
$(p\mu^-)p$ and $(d\mu^-)d$ results correspond to the ones given
in \cite{kilic04}. Obviously, the same results apply,
respectively, for $(p\mu^-)\bar{p}$ and $(d\mu^-)\bar{d}$. Table I
displays the striking difference of such results when compared
with the ones for $(p\bar{p})\mu^-$ and $(d\bar{d})\mu^-$. Related
to $Ps^-$ we should note that our approach gives $\lambda=8$, to
be compared with $\lambda=7.06$ obtained by Botero~\cite{botero86}
in a more involved calculation.

\begin{table}\vspace{-0.3cm}
\caption{
Strength $\lambda$  of the attractive dipole interaction (in dimensionless
units) generated by the charges $q_1$ and $q_2=-q_1$, when capturing a charge $q_3$,
for $|q_3|=|q_1|$ and different mass configurations. In the examples, the
estimated value for $\lambda$ is the same for the cases with the corresponding
antiparticle configurations.
Estimations of level ratios
${\cal {E}}_{\nu-1}/{\cal {E}}_\nu $ are given when $\lambda>1/4$.
The masses $m$ and $M$ are for arbitrary defined particles, with condition $M>>m)$.
}
{\begin{tabular}{@{}lcc|lcc@{}}
\hline\hline
\multicolumn{6}{c}{}\\[-2ex]
System&$\lambda$&${\cal {E}}_{\nu-1}/{\cal {E}}_\nu$& System&$\lambda$&
${\cal {E}}_{\nu-1}/{\cal {E}}_\nu $\\
\hline\hline
$(p\bar{p})e^\pm$     & 0.0065      & ---&
$(\mu^+\mu^-)e^\pm$   & 0.058       & ---\\
$(K^+K^-)e^\pm$       & 0.012       & ---&
$(p\mu^-)e^\pm$       &  0.025      & ---\\
$(m^-m^+)m^\pm$       &    8        & 9.55&
$(M^\pm m^\mp)m^\mp$  &    6        & 13.74\\
$(e^-e^+)p$           &    24       & 3.63&
$(e^-e^+)\mu^\pm$     & 23.75       & 3.66\\
$(p\mu^-)\mu^\pm$     &  6.06       & 13.55&
$(pK^-)K^\pm$         &  6.81       & 11.63\\
$(pe^-)p$             & 5508        & 1.09&
$(pe^-)\mu^\pm$       & 1428        & 1.18\\
$(p\bar{p})\pi^\pm$   &  1.66       & 198.62&
$(p\bar{p})\mu^\pm$   &  1.28       & 488.34\\
$(p\bar{p})K^\pm$     &    5        & 17.87&
$(p\bar{p})d$         &   12        & 6.25\\
$(d{\bar d})\mu^\pm$  & 0.66        & 1.8$\times10^{4}$&
$(t{\bar t})\mu^\pm$  & 0.44        & 1.8$\times10^{6}$\\
$(p\pi^-)p$           & 24.77       & 3.56&
$(p\mu^-)p$           & 31.22       & 3.09\\
$(d\pi^-)d$           & 44.89       & 2.56&
$(d\mu^-)d$           & 57.82       & 2.29\\
$(t\pi^-)t$           & 65.04       & 2.18&
$(t\mu^-)t$           & 84.45       & 1.98\\
$(t\pi^-)p$           & 32.13       & 3.04&
$(t\mu^-)p$           & 41.84       & 2.65\\
$(p\pi^-)t$           & 38.49       & 2.76&
$(p\mu^-)t$           & 48.11       & 2.48\\
\hline\hline
\end{tabular}}
\vspace{-.5cm}
\end{table}

\section{Discussion and summary}

{Before our conclusions, it is relevant a discussion
on the appli\-cabi\-lity of the Born-\-Oppenheimer ap\-pro\-xima\-tion
to the particular con\-figuration of excited states in three-charged
systems.
The separation of variables appro\-ximation, expressed in the
assumption of a product form for the wave function of the
three-body system, considered in Eq.~(\ref{eq04}),
breaks down when the third particle gets close to the neutral subsystem,
with $R$ of the order of the corresponding Bohr radius.
For some three-body systems, as for example systems with proton and
antiproton, nuclear force effects can also be relevant to be considered
in addition to Coulomb effects.
While in the asymptotic region the excited-state three-charged wave function
can be described well by the Bohr-Oppenheimer approximation, as
the size of the system gets smaller the magnitude of non-adiabatic
effects is enhanced.
In this case, in order to go beyond the adiabatic approximation,
the appropriate description of the wave function will consist of an
expansion with several higher-order terms, such that one cannot
disregard the coupling between them.
However, once one large three-charged state is formed with a size
much larger than the dipole radius, i.e., with the wave function
having major contribution from the asymptotic region, the
non-adiabatic effects can be translated to a boundary condition on
the wave function, at a radius characteristic of the region where
non-adiabatic effects start to be important, which can be roughly
estimated to be of the order of few Bohr radii.
Within our present approximation we cannot obtain the first large
bound excited state that spreads out in the asymptotic region of
the dipole potential, as its energy is determined by non-adiabatic
contributions.
But, once this energy is known, the energies of the other large
excited states follow the expression derived from the dipole interaction,
emerging the geometrically scaled pattern given in Eq.~(\ref{eq12}).
}

In summary, considering a charged particle captured by a dimer
(bound two-body subsystem), within an adiabatic approximation to
the Coulombic three-body system with arbitrary masses, we derived
the strength of the attractive dipole interaction with the
critical condition for level condensation below the first excited
state of the dimer, as given in Eqs.~(\ref{eq10}) and
(\ref{eq11}). The adiabatic approach is justified when the
energy of the relative motion of the dimer center-of-mass, with
respect to the third particle, is small compared to the subsystem
binding energy; and physically reasonable for a weak dipole
potential interacting with the spectator particle. The particular
straightforward charge-mass expression obtained for the leading
term of the dipole interaction, as well as the condition to occur
the {geometrically scaled energy spectrum}, can be very useful
as an insight into the analysis of three-body charged systems with
different mass relations. Of particular interest is the use of
such approach to study interaction properties of matter and
antimatter. Another relevant application can be found in the
interaction of excitons with electrons in
semiconductors~\cite{butov02}.
 As the electron and hole in the interacting pair of excitons can acquire effective individual
 masses distinct from the free electron mass~\cite{tan2005}, a general charge-mass conditions
 can be very helpful to analyze the relation between effective masses and the observed spectrum.
We finally note that the present approach can be very useful to study capture reactions of a
charged particle by a neutral two-body system, in view of possible three-body configurations.
From a configuration without spectrum below the state $n=2$, one can generate a system with
very dense spectrum by exchanging the spectator particle with the same charge particle
of the neutral system. An exchange mechanism can change drastically the observed spectrum: for
$(e^-e^+)p$ we have  ${\cal {E}}_{\nu-1}/{\cal {E}}_\nu=$3.63 (levels below the first excited
state of $Ps$), whereas for $(e^-p)e^+$ we have  ${\cal {E}}_{\nu-1}/{\cal {E}}_\nu=$13.74
(levels below the first excited state of the Hydrogen atom).

LT thanks Profs. S.B. Duarte and R. Shellard for the hospitality in CBPF.
We acknowledge financial support from Funda\c c\~ao de Amparo \`a Pesquisa do Estado de
S\~ao Paulo and Conselho Nacional de Desenvolvimento Cient\'\i fico e Tecnol\'ogico.

\end{document}